\documentclass[prl,twocolumn,showpacs]{revtex4}
\usepackage{graphicx}
\begin{document}

\renewcommand{\b}[1]{\mathbf{#1}}
\renewcommand{\c}[1]{\mathcal{#1}}
\newcommand{\expc}{\mathop{\mathrm{expc}}}
\renewcommand{\Im}{\mathop{\mathrm{Im}}}
\newcommand{\Tr}{\mathop{\mathrm{Tr}}}
\newcommand{\Res}{\mathop{\mathrm{Res}}}

\title{Nonequilibrium current driven by a step voltage pulse: an exact solution }
\author{Joseph Maciejko$^1$, Jian Wang$^2$ and Hong Guo$^1$}
\affiliation{1. Center for the Physics of Materials and Department
of Physics, McGill University, Montreal, Quebec, Canada, H3A 2T8\\
2. Department of Physics, The University of Hong Kong, Pok Fulam Road, Hong
Kong }
\date{\today}

\begin{abstract}
One of the most important problems in nanoelectronic device theory is to estimate how fast or how slow a quantum device can turn on/off a current. For an arbitrary noninteracting phase-coherent device scattering region connected to the outside world by leads, we have derived an exact solution for the nonequilibrium, nonlinear, and time-dependent current driven by both up- and down-step pulsed voltages. Our analysis is based on the Keldysh nonequilibrium Green's functions formalism where the electronic structure of the leads as well as the scattering region are treated on an equal footing. A model calculation for a quantum dot with a Lorentzian linewidth function shows that the time-dependent current dynamics display interesting finite-bandwidth effects not captured by the commonly used wideband approximation.
\end{abstract}


\maketitle

Understanding coherent transport of charge and spin through the
scattering region of a nanoscale device is the central problem of
nanoelectronics theory which has tremendous scientific and
technological importance \cite{roadmap}. From both theory and
application points of view, an important issue which has yet to be
resolved is to predict how fast or how slow a nanoelectronic device
can turn on/off a current from quantum mechanical first principles.
Indeed, one cannot develop an electronic technology unless the
operational speed of the device can be designed and controlled. This
issue is closely related to the transient transport phenomenon,
which is becoming an extremely important problem of nanoelectronic
device physics, as can be observed in such effects as
photon-assisted tunneling \cite{Kouwenhoven1994}, electron
turnstiles \cite{Kouwenhoven1991} and ringing behavior in the
time-dependent current \cite{Wingreen1993,Jauho1994}. Recent
real-time measurements of electron dynamics \cite{Lu2003} have
further raised interest for the study of transient quantum
transport.

The purpose of this work is to investigate transient quantum
transport far from equilibrium, for nanoelectronic systems in the
Lead-Device-Lead (LDL) configuration where ``Device'' indicates the
scattering region which is connected to the outside world by the
leads. When the LDL system is driven far from equilibrium by a
step-shaped voltage pulse, we discovered an \emph{exact} solution to
the time-dependent current $J(t)$, thereby an \emph{exact} transient
quantum transport picture is obtained. To the best of our knowledge,
this is the first time an exact solution is found for
far-from-equilibrium transient transport dynamics in the quantum
regime, and it provides a valuable and unambiguous physical picture
of how charge current is turned on and off by a bias voltage pulse
through devices of the LDL form. Importantly, the exact solution
provides, for example, the correct current decay time scale after
the bias voltage is turned off, and the transient current that
follows the turning-on of a constant voltage.

For quantum devices in the form of LDL, the theoretical formalism
best suited to the study of time-dependent transport is, perhaps,
the Keldysh nonequilibrium Green's
functions \cite{Wingreen1993,Jauho1994}. In the NEGF formalism, the
time-dependent current through the phase-coherent scattering region
of the device is given in terms of local Green's functions. However,
these Green's functions cannot, in general, be solved analytically.
Previous studies of time-dependent transport have thus so far relied on
the so-called wideband limit (WBL) \cite{Wingreen1989}, which is a
simplifying assumption where the coupling between the scattering
region and the external leads is taken to be independent of energy.
In other words, the WBL neglects any electronic structure of the device
leads. Therefore, when the electronic structure of the leads is
important, {\it i.e.} for leads with finite bandwidth such as those made
of semiconductors, nanotubes, nanowires, etc., a theory beyond the WBL
is necessary.  In this regard, \emph{numerical }approaches have been
put forward in Refs. \onlinecite{Kwapinski2002,Zhu2005}.

Our starting point is the formalism described in
Refs. \onlinecite{Wingreen1993,Jauho1994,HaugJauho}. The Hamiltonian
of the LDL device is
\begin{eqnarray}
\label{hamiltonian}
H&=&\sum_{\b{k}\alpha}\epsilon_{\b{k}\alpha}(t)c_{\b{k}\alpha}^\dag c_{\b{k}\alpha}+
\sum_{mn}\epsilon_{mn}(t) d_m^\dag d_n\nonumber\\
&&+\sum_{\mathbf{k}\alpha,n}\left(t_{\mathbf{k}\alpha,n}c_{\mathbf{k}\alpha}^\dag d_n+
t_{\mathbf{k}\alpha,n}^*d_n^\dag c_{\mathbf{k}\alpha}\right),
\end{eqnarray}
where $c_{\b{k}\alpha}^\dag$ ($c_{\b{k}\alpha}$) with $\alpha=L,R$
creates (destroys) an electron with momentum $\b{k}$ in the left
($L$) or right ($R$) lead, and $d_n^\dag$ ($d_n$) creates (destroys)
an electron in a single-particle state labeled by $n$ in the
scattering region. Quantities $t_{\mathbf{k}\alpha,n}$ describe
coupling of the leads to the scattering region of the device.
Chemical potentials in both leads are set to zero. As in Refs.
\onlinecite{Wingreen1993,Jauho1994,HaugJauho}, when an external
time-dependent voltage is applied to drive a current through the
device, we assume that the single-particle energies acquire a
time-dependent shift:
$\epsilon_{\b{k}\alpha}(t)=\epsilon_{\b{k}\alpha}^0+\Delta_\alpha(t)$
and $\epsilon_{mn}(t)=\epsilon_{mn}^0+\Delta_{mn}(t)$. It has been
shown \cite{Jauho1994} that the charge current through lead $\alpha$
is given by
\begin{eqnarray}\label{jauho}
J_\alpha(t)&=&-2e\int_{-\infty}^t dt'\int\frac{d\epsilon}{2\pi}\Im\Tr\{e^{i\epsilon(t-t')}
e^{i\int_{t'}^t dt_1\,\Delta_\alpha(t_1)}\nonumber\\
&&\times\Gamma_\alpha(\epsilon)[G^<(t,t')+f(\epsilon)G^R(t,t')]\},
\end{eqnarray}
where $f(\epsilon)\equiv(e^{\beta\epsilon}+1)^{-1}$ is the Fermi
function, the linewidth function $\Gamma_\alpha(\epsilon)$ has
matrix elements
$\Gamma_{\alpha,mn}(\epsilon)\equiv2\pi\rho_\alpha(\epsilon)
t_{\alpha,m}^*(\epsilon)t_{\alpha,n}(\epsilon)$ where
$\rho_\alpha(\epsilon)$ is the density of states in lead $\alpha$,
the lesser Green's function $G^<(t,t')$ has matrix elements
$G^<_{nm}(t,t')\equiv i\langle d_m^\dag(t')d_n(t)\rangle$ and the
retarded Green's function $G^R(t,t')$ has matrix elements
$G^R_{nm}(t,t')\equiv-i\theta(t-t')\langle\{d_n(t),d_m^\dag(t')\}\rangle$.
The retarded Green's function is given by the solution of the Dyson
equation \cite{Mahan,HaugJauho} $G^R(t,t')=G_0^R(t,t')+\int dt_1\int
dt_2\,G_0^R(t,t_1)\Sigma^R(t_1,t_2)G^R(t_2,t')$, where $G_0^R(t,t')$
is the Green's function of the scattering region without any leads,
while the Keldysh equation yields the lesser Green's function
$G^<(t,t')=\int dt_1\int
dt_2\,G^R(t,t_1)\Sigma^<(t_1,t_2)G^A(t_2,t')$, where $G^A(t,t')$ is
the advanced Green's function and $\Sigma^{R,<}(t_1,t_2)$ are the
retarded and lesser self-energies, respectively. From a mathematical
point of view, the main effect of broken time-translational
invariance due to the presence of time-dependent external fields is
that the double integral Dyson equation is not a simple Fourier
convolution product as is the case in equilibrium or for
steady-state transport, and therefore cannot be reduced by a Fourier
transformation to a simple algebraic matrix equation. In the WBL, after
one neglects the energy dependence of the linewidth function, the
self-energy $\Sigma^R(t_1,t_2)$ becomes proportional to a delta
function $\delta(t_1-t_2)$, hence $G^R(t,t')$ can be solved
afterward \cite{Wingreen1989,Wingreen1993,Jauho1994}.

To solve the problem exactly without relying on the WBL, we
define \cite{Wingreen1993,Jauho1994}
\begin{equation}\label{defA}
A_\alpha(\epsilon,t)\equiv \int_{-\infty}^t dt'\,e^{i\epsilon(t-t')}
e^{i\int_{t'}^t dt_1\,\Delta_\alpha(t_1)}G^R(t,t').
\end{equation}
From the Keldysh equation and the expression for the lesser
self-energy \cite{Jauho1994},
\[ \Sigma^<(t_1,t_2)=\sum_\alpha\int\frac{d\omega}{2\pi}\,e^{-i\omega(t_1-t_2)}
e^{i\int_{t_1}^{t_2}dt_3\,\Delta_\alpha(t_3)}f(\omega)\Gamma_\alpha(\omega),
\]
one can straightforwardly show that Eq. (\ref{jauho}) takes the form
\begin{equation}
J_\alpha(t)=-2e\int\frac{d\epsilon}{2\pi}\Im\Tr\{\Gamma_\alpha(\epsilon)
[\Psi_\alpha(\epsilon,t)+f(\epsilon)A_\alpha(\epsilon,t)]\},
\label{curr1}
\end{equation}
where we have defined
\begin{eqnarray}
& &\Psi_\alpha(\epsilon,t)\equiv i\sum_\beta\int\frac{d\epsilon'}{2\pi}\,
e^{i(\epsilon-\epsilon')t}f(\epsilon')A_\beta(\epsilon',t)\Gamma_\beta(\epsilon')\nonumber\\
&&\times\int_{-\infty}^t
dt'\,e^{-i(\epsilon-\epsilon')t'}e^{i\int_{t'}^t
dt_1[\Delta_\alpha(t_1)-\Delta_\beta(t_1)]}A_\beta^\dag(\epsilon',t')\ .\nonumber
\end{eqnarray}
Therefore, once $A_\alpha(\epsilon,t)$ can be found, the current is
entirely determined.

To solve for $A_\alpha(\epsilon,t)$, we make use of an alternate
version of the Dyson equation, namely that obtained by choosing only
the time-dependent part of the Hamiltonian (\ref{hamiltonian}) as a perturbation, the unperturbed part being the
time-independent terms. This way, the unperturbed Hamiltonian
describes the LDL device at \emph{equilibrium}, hence its physics is
given by a time-translationally invariant Green's function
$\tilde{G}^R(t-t')$. A standard expansion of the contour $S$-matrix
along the Schwinger-Keldysh contour
$C$ \cite{Schwinger1961,Keldysh1965}, followed by analytic
continuation of the contour-ordered Green's function
$G_{nm}(\tau,\tau')\equiv-i\langle
T_C\{d_n(\tau)d_m^\dag(\tau')\}\rangle$ to the real time axis
according to the Langreth rules \cite{Langreth1976}, yields the
following alternate form of the Dyson equation,
\begin{eqnarray}
& & G^R(t,t')=\tilde{G}^R(t-t')+\int dt_1\,\tilde{G}^R(t-t_1)\Delta(t_1)G^R(t_1,t')\nonumber\\
&&+\int dt_1\int dt_2\,\tilde{G}^R(t-t_1)V^R(t_1,t_2)G^R(t_2,t'),
\label{dyson1}
\end{eqnarray}
where we define a retarded potential \cite{Zhu2005}
$V^R(t_1,t_2)\equiv\sum_\beta\left[\exp\left(-i\int_{t_2}^{t_1}dt'\,
\Delta_\beta(t')\right)-1\right]\tilde{\Sigma}^R_\beta(t_1-t_2)$ and
$\tilde{\Sigma}^R_\beta$ is the equilibrium retarded self-energy due
to lead $\beta$. Using Eq. (\ref{defA}) in the Dyson equation
(\ref{dyson1}), we obtain the following integral equation for the
quantity $A_\alpha(\epsilon,t)$:
\begin{eqnarray}\label{eqA}
A_\alpha(\epsilon,t)&=&\tilde{A}_\alpha(\epsilon,t)+\int dt'\,e^{i\epsilon(t-t')}
e^{i\int_{t'}^t dt_1\,\Delta_\alpha(t_1)}\tilde{G}^R(t-t')\nonumber\\
&&\times\Delta(t')A_\alpha(\epsilon,t')+\int dt_1\int dt_2\,e^{i\epsilon(t-t_2)}\\
&&\times e^{i\int_{t_2}^t
dt_3\,\Delta_\alpha(t_3)}\tilde{G}^R(t-t_1)
V^R(t_1,t_2)A_\alpha(\epsilon,t_2).\nonumber
\end{eqnarray}
Here $\tilde{A}_\alpha(\epsilon,t)\equiv\int_{-\infty}^t
dt'\,e^{i\epsilon(t-t')}e^{i\int_{t'}^t
dt_1\,\Delta_\alpha(t_1)}\tilde{G}^R(t-t')$ describes the
equilibrium state and is therefore known. To solve Eq. (\ref{eqA}),
we need to specify the external fields $\Delta(t)$ and
$\Delta_\alpha(t)$. Here we investigate a step function pulse
applied at time $t=0$, of the form
$\Delta_{(\alpha)}(t)=\Delta_{(\alpha)}\theta(\pm t)$ where
$\Delta_{(\alpha)}$ is a constant amplitude and the plus (minus)
sign corresponds to an upward (downward) step.

\noindent \textbf{Downward step pulse.} This is the situation where
a constant bias voltage with value $\Delta_\alpha$ is sharply turned
off at time $t=0$ and remains off for subsequent times. For this
case and from its definition, $V^R(t_1,t_2)$ vanishes when $t_1$ and
$t_2$ are simultaneously greater than zero, as well as when
$t_1<t_2$ from the retarded self-energy. Equation (\ref{eqA}) then
takes the form
\begin{widetext}
\begin{eqnarray}\label{eqAdown}
A_\alpha(\epsilon,t)&=&\tilde{A}_\alpha(\epsilon,t)+\int_{-\infty}^0 dt'\,
e^{i\epsilon(t-t')}e^{i\int_{t'}^t dt_1\,\Delta_\alpha(t_1)}\tilde{G}^R(t-t')
\Delta A_\alpha(\epsilon,t')\nonumber\\
&&+\left(\int_{-\infty}^0 dt_1\int_{-\infty}^{t_1}dt_2+\int_0^t dt_1
\int_{-\infty}^0 dt_2\right)e^{i\epsilon(t-t_2)}e^{i\int_{t_2}^t dt_3\,\Delta_\alpha(t_3)}
\tilde{G}^R(t-t_1)V^R(t_1,t_2)A_\alpha(\epsilon,t_2).
\end{eqnarray}
\end{widetext}
We see from the limits of integration that while
$A_\alpha(\epsilon,t)$ is required for $t>0$ on the left-hand side
of Eq. (\ref{eqAdown}), only $A_\alpha(\epsilon,t<0)$ is involved in
the integrals on the right-hand side. This is an example of a
Wiener-Hopf equation \cite{MorseFeshbach}. For our LDL device under
this downward step pulse, $A_\alpha(\epsilon,t<0)$ is actually
known: for $t<0$ the system is in steady-state under a constant bias
$\Delta_\alpha$. Hence $A_\alpha(\epsilon,t<0)$ is easily obtained
from the known steady-state NEGF $\bar{G}^R(t-t')\equiv
G^R(t<0,t'<0)$. Equation (\ref{eqAdown}) is therefore \emph{not} an integral
equation but an explicit expression for $A_\alpha(\epsilon,t)$ in
terms of known quantities. Once the integrations are carried out by
tedious but elementary algebra, one obtains the following exact
expression:
\begin{eqnarray}
\label{Adown}
A_\alpha(\epsilon,t)&=&\tilde{G}^R(\epsilon)+\int\frac{d\omega}{2\pi i}\,
\frac{e^{-i(\omega-\epsilon)t}\tilde{G}^R(\omega)}{\omega-\epsilon-\Delta_\alpha-i0^+}\left[\frac{\Delta_\alpha}{\omega-\epsilon-i0^+}\right.\nonumber\\
&&\left.+\left(\Delta-\sum_\beta\Delta_\beta\tilde{\Upsilon}^R_{\alpha\beta}(\omega,\epsilon)\right)
\bar{G}^R(\epsilon+\Delta_\alpha)\right],
\end{eqnarray}
where we define
$\tilde{\Upsilon}^R_{\alpha\beta}(\omega,\omega')\equiv
[\tilde{\Sigma}^R_\beta(\omega)-\tilde{\Sigma}^R_\beta(\omega'+\Delta_\alpha-\Delta_\beta)]/
(\omega-\omega'-\Delta_\alpha+\Delta_\beta)$. Equation (\ref{Adown}) is the
first important result of this paper. With the explicit solution for
$A_\alpha(\epsilon,t)$, the time-dependent current $J_\alpha (t)$
can be obtained without further difficulty from Eq. (\ref{curr1}).

\noindent \textbf{Upward step pulse.} In this case, Eq. (\ref{eqA})
takes the form
%\begin{widetext}
\begin{eqnarray}\label{eqAup}
&& A_\alpha(\epsilon,t)=A'_\alpha(\epsilon,t)+\int_0^t dt'\,e^{i(\epsilon+\Delta_\alpha)(t-t')}\tilde{G}^R(t-t')\nonumber\\
&&\times\Delta A_\alpha(\epsilon,t')+\int_0^t dt_1\int_0^{t_1}dt_2\,e^{i(\epsilon+\Delta_\alpha)(t-t_1)}\nonumber\\
&&\times\tilde{G}^R(t-t_1)e^{i(\epsilon+\Delta_\alpha)(t_1-t_2)}V^R(t_1-t_2)A_\alpha(\epsilon,t_2),\nonumber
\end{eqnarray}
%\end{widetext}
where $A'_\alpha(\epsilon,t)$ is a known function that involves only
$\tilde{A}_\alpha(\epsilon,t)$ and $A_\alpha(\epsilon,t<0)$. This is
a Volterra equation which has the form of a Laplace convolution
product. It can thus be converted into an algebraic matrix equation
by a Laplace transformation, so that $A_\alpha(\epsilon,t)$ is given
by the following Bromwich integral:

\begin{eqnarray}
A_\alpha(\epsilon,t)&=&\frac{1}{2\pi i}\int_{\gamma-i\infty}^{\gamma+i\infty}d\sigma\,e^{\sigma t}\nonumber\\
&&\times\left(1-F_\alpha(\epsilon,\sigma)[\Delta+U_\alpha(\epsilon,\sigma)]\right)^{-1}
A'_\alpha(\epsilon,\sigma),\nonumber
\end{eqnarray}
where $\sigma$ is the Laplace variable, and we define
$A'_\alpha(\epsilon,\sigma)\equiv
\mathcal{L}_{t\rightarrow\sigma}\{A'_\alpha(\epsilon,t)\}$,
$F_\alpha(\epsilon,\sigma)\equiv
\mathcal{L}_{t\rightarrow\sigma}\{e^{i(\epsilon+\Delta_\alpha)t}\tilde{G}^R(t)\}$
and $U_\alpha(\epsilon,\sigma)\equiv
\mathcal{L}_{t\rightarrow\sigma}\{e^{i(\epsilon+\Delta_\alpha)t}V^R(t)\}$
which are all known quantities. Finally, a detailed but straightforward calculation gives
\begin{eqnarray}
\label{Aup}
&&A_\alpha(\epsilon,t)=e^{i\Delta_\alpha t}\tilde{G}^R(\epsilon)+
\int_{C_\gamma}\frac{dz}{2\pi i}\frac{e^{-izt}\bar{G}^R(\epsilon+z+\Delta_\alpha)}{z+\Delta_\alpha}\nonumber\\
&&\times\left[\frac{\Delta_\alpha}{z}+
\left(\Delta-\sum_\beta\Delta_\beta\tilde{\Upsilon}^R_{\alpha\beta}(\epsilon,\epsilon+z)\right)
\tilde{G}^R(\epsilon)\right],
\end{eqnarray}
where we define the contour
$C_\gamma:\infty+i\gamma\rightarrow-\infty+i\gamma$ in the complex
$z$-plane. Equation (\ref{Aup}) is the second important result of this
work.

\noindent \textbf{Examples.} The above exact solutions, Eqs.
(\ref{Adown},\ref{Aup}), are valid for arbitrary noninteracting
phase-coherent devices in the LDL configuration. As an concrete
example, we now apply these results to a quantum dot (QD) with a
single energy level $\epsilon_0$ connected to two leads described by
a Lorentzian linewidth function
$\Gamma_\alpha(\epsilon)=\Gamma_\alpha^0W^2/(\epsilon^2+W^2)$, where
$\Gamma_\alpha^0$ is a constant linewidth amplitude and $W$ is the
bandwidth. After the integrals in Eqs. (\ref{Adown},\ref{Aup}) are
carried out by residue integration, the time-dependent current
$J_\alpha(t)$ of Eq. (\ref{curr1}) is then given by a single
integral over all frequencies. This last integral can be easily done
numerically as it contains $f(\epsilon)$ which has known poles at
the fermionic Matsubara frequencies $i\omega_n=i(2n+1)\pi/\beta$.

\begin{figure}
\includegraphics[scale=0.62]{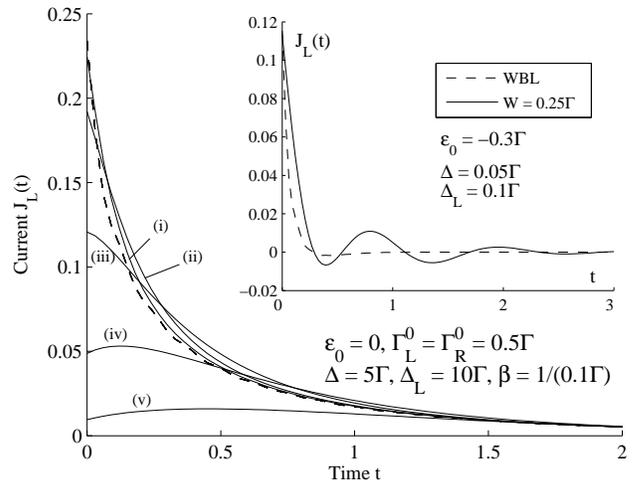}
\caption{Time-dependent current $J_L(t)$ through left lead in
response to a downward step pulse for different bandwidths: dashed
line, WBL ($W=\infty$); (i) $W=20\Gamma$, (ii) $W=10\Gamma$, (iii)
$W=5\Gamma$, (iv) $W=2.5\Gamma$, and (v) $W=\Gamma$. The current is
in units of $e\Gamma/\hbar$ and the time is in units of
$\hbar/\Gamma$ where $\Gamma=\Gamma_L^0+\Gamma_R^0$ is the total
linewidth amplitude. Parameters are taken the same as in Ref.
\onlinecite{Wingreen1993}. When $W=100\Gamma$, the result was found
(not shown) to be indistinguishible from the $W=\infty$ curve. The
inset shows interesting oscillatory behavior, not captured by the
WBL, for a narrow band $W=0.25\Gamma$ misaligned with the resonant
level
 $\epsilon_0=-0.3\Gamma$.} \label{fig:downbw}
\end{figure}

In Fig. \ref{fig:downbw} we show the time-dependent current $J_L(t)$
through the left lead in response to a \emph{downward} step pulse.
At $t=0$, energies in the left lead are lowered by
$\Delta_L=10\Gamma$ and the energy of the QD level is lowered by
$\Delta=5\Gamma$ where $\Gamma$ is the total linewidth amplitude.
For $W\gg\Gamma,\Delta_L,\Delta$, the WBL result is essentially
correct. When $W$ becomes comparable to other energy scales of the
problem, the WBL is seen to be a poor approximation. The initial
current $J_L(0)$ decreases as the linewidth function gets narrower
since less states in the leads are available for transport. More
importantly, the time-dependent current can increase following the
bias turnoff (curves (iv) and (v) in Fig. \ref{fig:downbw}), an
interesting nonclassical behavior not displayed by the WBL current.
A current increase after the bias is turned off was also observed in
a previous numerical study \cite{Zhu2005}. Most importantly, for
devices with smaller bandwidth, the WBL and the exact solution
predict very different time scales of the current decay. The inset
shows an interesting oscillatory behavior, not captured by the WBL,
for a narrow band $W=0.25\Gamma$ misaligned with the resonant level
$\epsilon_0=-0.3\Gamma$.

\begin{figure}
\includegraphics[scale=0.62]{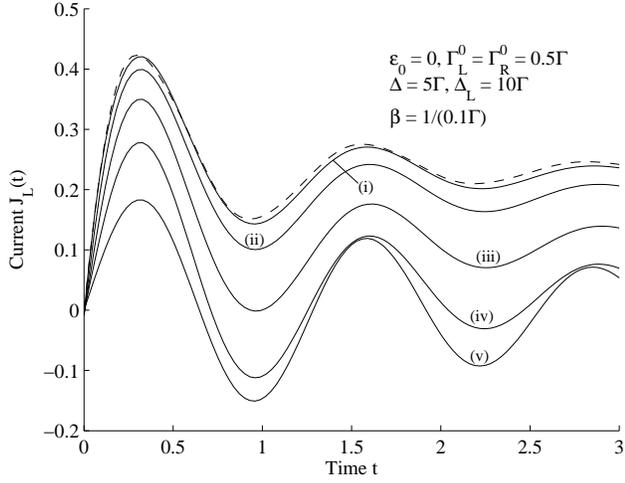}
\caption{Time-dependent current $J_L(t)$ through left lead in
response to an upward step pulse for different bandwidths: dashed
line, WBL ($W=\infty$); (i) $W=20\Gamma$, (ii) $W=10\Gamma$, (iii)
$W=5\Gamma$, (iv) $W=2.5\Gamma$, and (v) $W=\Gamma$. Units and
parameters are the same as in Fig. \ref{fig:downbw}. Here again, the
$W=100\Gamma$ curve (not shown) is indistinguishible from the
$W=\infty$ curve.} \label{fig:upbw}
\end{figure}

In Fig. \ref{fig:upbw} we show the time-dependent current $J_L(t)$
through the left lead in response to an \emph{upward} step pulse. In
this case, the single-particle energies $\epsilon_{\b{k}\alpha}$ and
$\epsilon_0$ are suddenly raised at $t=0$. Here again, the WBL is
seen to be accurate for $W\gg\Gamma,\Delta_L,\Delta$. However,
interesting finite-bandwidth effects appear as bands in the leads
get narrower. First of all, the asymptotic $t\rightarrow\infty$
current decreases with $W$, corresponding to the initial current
decrease in the downward step situation discussed in the last
paragraph. In addition, it is seen that a \emph{positive} voltage
pulse can drive an instantaneously \emph{negative }current, as has
been observed in the numerical work of Ref. \cite{Zhu2005}. For
$W<\Delta,\Delta_L$, the current oscillates around a zero value: the
pulse drives the resonant level $\epsilon_0$ outside the band so
that little current can flow through.

In summary, we have presented an exact solution for the
time-dependent current through an arbitrary noninteracting
phase-coherent device scattering region connected to external leads
with arbitrary energy-dependent linewidth functions, in the
physically relevant case of an upward or downward step function
voltage pulse. The results are general and are valid for far from
equilibrium transport situations. For a single-level QD with
Lorentzian linewidth function, the WBL was seen to be a crude
approximation to the exact solution in the case of narrow bands. Due
to the finite-bandwidth effects, a number of nonclassical transient
behaviors were found including a current increase following a
downward pulse, a negative current driven by a positive upward
pulse, and a vanishing asymptotic current for a finite positive bias
pulse. The significance of our solution is the exactness of the
results which give unambiguous nonequilibrium transient quantum
transport dynamics. From a practical application point of view, our
formalism gives the transient current in terms of steady-state NEGF
that can be calculated by any technique used for steady-state
transport such as those atomistic first principles techniques of
Ref. \onlinecite{mcdcal}.

\noindent {\bf Acknowledgments.} We thank Dr. Eric Zhu, Mr. Tao Ji
and Mr. Derek Waldron for discussions concerning the NEGF theory and
numerical issues. We gratefully acknowledge financial support from
NSERC (H.G., J.M.); CIAR (H.G.); Richard H. Tomlinson Fellowship
(J.M.); and a RGC grant from the SAR Government of Hong Kong under
grant number HKU 7044/05P.
%\end{acknowledgments}

%\bibliography{paper.bib}

\begin{thebibliography}{00}

\bibitem{roadmap}
{\it International Technology Roadmap for Semiconductors,}
Semiconductor Industry Association (2004).

\bibitem{Kouwenhoven1994}
L.P. Kouwenhoven {\it et al.}, Phys. Rev. Lett. {\bf 73}, 3443
(1994).

\bibitem{Kouwenhoven1991}
L.P. Kouwenhoven {\it et al.}, Phys. Rev. Lett. {\bf 67}, 1626
(1991).

\bibitem{Wingreen1993}
N.S. Wingreen, A.-P. Jauho and Y. Meir, Phys. Rev. B {\bf 48}
8487 (1993).

\bibitem{Jauho1994}
A.-P. Jauho, N.S. Wingreen and Y. Meir, Phys. Rev. B
\textbf{50}, 5528 (1994).

\bibitem{Lu2003}
W. Lu {\it et al.}, Nature {\bf 423}, 422 (2003).

\bibitem{Wingreen1989}
N.S. Wingreen, K.W. Jacobsen and J.W. Wilkins, Phys. Rev. B {\bf
40}, 11834 (1989).

\bibitem{Kwapinski2002}
T. Kwapi\'{n}ski, R. Taranko and E. Taranko, Phys. Rev. B {\bf 66},
035315 (2002).

\bibitem{Zhu2005}
Y. Zhu {\it et al.}, Phys. Rev. B {\bf 71}, 075317 (2005).

\bibitem{HaugJauho}
H. Haug and A.-P. Jauho, {\it Quantum Kinetics in Transport and
Optics of Semiconductors}, (Springer-Verlag, Berlin, 1998).

\bibitem{Mahan}
G.D. Mahan, {\it Many-Particle Physics}, (Kluwer Academic, New
York, 2000).

\bibitem{Schwinger1961}
J. Schwinger, J. Math. Phys. \textbf{2}, 407 (1961).

\bibitem{Keldysh1965}
L.V. Keldysh, Zh. Eksp. Teor. Fiz. {\bf 47}, 1515 (1964) [Sov. Phys. JETP {\bf 20}, 1018 (1965)].

\bibitem{Langreth1976}
D.C. Langreth, ``Linear and nonlinear response theory with
applications", in {\it Linear and Nonlinear Electron Transport in
Solids}, NATO Advanced Study Institute Series, vol. 17, (Plenum
Press, 1976).

\bibitem{MorseFeshbach}
P.M. Morse and H. Feshbach, {\it Methods of Theoretical Physics},
Vol. I (McGraw-Hill, New York, 1953).

\bibitem{mcdcal}
J. Taylor, H. Guo and J. Wang, Phys. Rev. B. {\bf 63}, 245407
(2001); {\it ibid}, {\bf 63}, 121104 (2001); Y. Xue, S. Datta and
M.A. Ratner, J. Chem. Phys. {\bf 115}, 4292 (2001); Chem. Phys. {\bf
281}, 151 (2002);
%M. Brandbyge, J.-L. Mozos, P. Ordej\'on, J. Taylor, and K. Stokbro,
M. Brandbyge {\it et al.}, Phys. Rev. B {\bf 65}, 165401 (2002);
J.J. Palacios {\it et al.}, Phys. Rev. B {\bf 66}, 035332 (2002);
S.-H. Ke, H.U. Baranger and W. Yang, Phys. Rev. B {\bf 70}, 085410
(2004); A.R. Rocha, {\it et al.}, Nature Materials {\bf 4}, 335
(2005).
\end{thebibliography}

\end{document}